\documentstyle[12pt]{article}
\newfont{\Mb}{msbm10}

\begin{document}
\hspace\parindent
\thispagestyle{empty}

\bigskip
\bigskip
\bigskip
\begin{center}
{\LARGE \bf Determining Liouvillian First Integrals for}
\end{center}
\begin{center}
{\LARGE \bf  Dynamical Systems in the Plane}
\end{center}

\bigskip

\begin{center}
{\large
$^{a,b}$J. Avellar, $^a$L.G.S. Duarte, $^{c}$ S.E.S. Duarte and $^a$L.A.C.P. da Mota \footnote{E-mails: javellar@dft.if.uerj.br, lduarte@dft.if.uerj.br, sduarte@dft.if.uerj.br and damota@dft.if.uerj.br}
}

\end{center}

\bigskip
\centerline{\it $^a$ Universidade do Estado do Rio de Janeiro,}
\centerline{\it Instituto de F\'{\i}sica, Depto. de F\'{\i}sica Te\'orica,}
\centerline{\it 20559-900 Rio de Janeiro -- RJ, Brazil}

\bigskip
\centerline{\it $^b$ Funda\c c\~ao de Apoio \`a Escola T\'ecnica,}
\centerline{\it E.T.E. Juscelino Kubitschek,}
\centerline{\it 21311-280 Rio de Janeiro -- RJ, Brazil}

\bigskip
\centerline{\it $^c$ C.F.E.T ,}
\centerline{\it 20537-200 Rio de Janeiro -- RJ, Brazil}

\bigskip
\bigskip
\abstract{Here we present/implement an algorithm to find
Liouvillian first integrals of dynamical systems in the plane. In
\cite{JCAM}, we have introduced the basis for the present
implementation. The particular form of such systems allows
reducing it to a single rational first order ordinary differential
equation (rational first order ODE). We present a set of software routines
in Maple 10 for solving rational first order ODEs. The package present
commands permitting research incursions of some algebraic
properties of the system that is being studied.

\bigskip
\bigskip
\bigskip



\centerline{ {\bf (Submitted to Computer Physics Communications)} }

\bigskip
\bigskip
\bigskip
\bigskip
\bigskip
\bigskip

\newpage
\bigskip
\hspace{1pc}
{\bf PROGRAM SUMMARY}
\bigskip

\begin{footnotesize}
\noindent
{\em Title of the software package:} {\it Lsolver}.   \\[10pt]
{\em Catalogue number:} (supplied by Elsevier)                \\[10pt]
{\em Software obtainable from:} CPC Program Library, Queen's
University of Belfast, N. Ireland.
\\[10pt]
{\em Licensing provisions:} none  \\[10pt]
{\em Operating systems under which the program has been tested:}
Windows ME, Windows XP.
\\[10pt]
{\em Programming languages used:} Maple 10
\\[10pt]
{\em Memory required to execute with typical data:}  128 Megabytes. \\[10pt]
{\em No. of lines in distributed program, including On-Line Help,
etc.:} 900.                                                   \\[10pt]
{\em Nature of mathematical problem}\\
Solution of rational first order ordinary differential equations.
\\[10pt]
{\em Methods of solution}\\
The method of solution is based on a Darboux/PS type approach.
\\[10pt]
{\em Restrictions concerning the complexity of the problem}\\
If the integrating factor for the rational first order ODE under consideration presents Darboux Polynomials of
high degree ( $> 3 $ ) in the dependent and independent variables, the package may spend an unpractical amount of time to obtain the solution.
\\[10pt]
{\em Typical running time}\\
This depends strongly on the ODE, but usually under 2 seconds.
\\[10pt]
{\em Unusual features of the program}\\
Our implementation not only search for Liouvillian conserved
quantities, but can also be used as a research tool that allows
the user to follow all the steps of the Darboux procedure (for
example, the algebraic invariants curves and associated cofactors
can be calculated). In addition, since our package is based in
recent theoretical developments \cite{JCAM}, it can successfully
solve a class of rational first order ODEs that were not solved by some of the
best-known ODE solvers available.
\end{footnotesize}
\newpage
\hspace{1pc}
{\bf LONG WRITE-UP}

\section{Introduction}
\label{intro}

The problem of searching for conserved quantities (first integrals) on
physical systems is not new. In the last thirty years there has been a
great increasing of interest in methods that follow the lines
drawn by Darboux \cite{Darboux} in the second half of $19^{th}$ century combined with topics
that comes from differential algebra (developed by Ritt \cite{Ritt} and
Kolchin\cite{Kol} in the middle of the $20^{th}$ century).

Particularly, the polynomial systems on the plane,
\begin{eqnarray}
\label{system}
\frac{dx}{dt}&=&N(x,y) \nonumber \\
\frac{dy}{dt}&=&M(x,y)
\end{eqnarray}
\noindent
where $M$ and $N$ are polynomials in $(x,y)$, have received a great deal of attention. We can then notice that there is an equivalence
in finding first integrals of this system with the solving of a rational first order
differential equation of the form:
\begin{equation}
\label{FOODE0}
y' = {\frac{dy}{dx}} = {\frac{M(x,y)}{N(x,y)}}.
\end{equation}
\noindent
If $I(x,y)$ is a first integral of the system (\ref{system}) then
$I(x,y)=K$, where $K$ is a constant, is a general solution of the first order
differential equation (\ref{FOODE0}).

A big step forward in constructing an
algorithm for solving first order ODEs analytically was taken in a
seminal paper by Prelle and Singer (PS)~\cite{PS} on autonomous systems of
ODEs. Prelle and Singer's problem is equivalent to asking when a rational first order ODE
of the form (\ref{FOODE0}) has an elementary solution (a solution which can be written
in terms of a combination of polynomials, logarithms, exponentials
and radicals).  Prelle and Singer were not exactly able to construct an
algorithm for solving their problem, since they were not able to define
a degree bound for the polynomials which might enter into the
solution. Though important from a theoretical point of view, any
computer implementation of the PS method will have a practical degree
bound imposed by the processing speed and/or memory needed to handle the
ever-more complex equations. With this in mind it is possible to say
that Prelle and Singer's original method is almost an algorithm, awaiting
a theoretical degree bound to turn it algorithmic. The approach of
Prelle and Singer \cite{PS} was so interesting that it has motivated
many extensions to its main ideas \cite{Shtokhamer,firsTHEOps1,secondTHEOps1,chris1,chris2,nossoPS2theo}

The purpose of this paper is to describe an implementation, in Maple, of
a Darboux-type procedure (presented on \cite{JCAM}) that extends the
applicability of the PS-approach to deal with rational first order ODEs (\ref{FOODE0}) with Liouvillian \footnote{For
a formal definition of Elementary and Liouvillian functions, please see
\cite{Davenport}.} solutions.

The paper is organized as follows: in section~\ref{PS}, we present a
short theoretical introduction to the PS approach; in the following
section, we introduce the main ideas of our extension \cite{JCAM}; in
section 4, we present examples of the application of our method; in section~\ref{package}, we present the
{\it Lsolver} package; section \ref{performance} talk about
some interesting features and performance of our package.

\section{The State of the art}
\label{PS}

Despite its usefulness in solving rational first order ODEs, the Prelle-Singer procedure is
not very well known outside mathematical circles, and so we present
a brief overview of the main ideas of the original PS-procedure and some
of its extensions.

\subsection{The Prelle-Singer Method}
\label{rational}

Consider the class of rational first order ODEs which can be written as
\begin{equation}
\label{FOODE}
y' = {\frac{dy}{dx}} = {\frac{M(x,y)}{N(x,y)}}
\end{equation}
where $M(x,y)$ and $N(x,y)$ are polynomials with coefficients in the complex
field $\it C$.

In~\cite{PS}, Prelle and Singer proved that, if an elementary first
integral of~(\ref{FOODE}) exists, it is possible to find an algebraic integrating
factor $R$ such that $R^n$ is a rational function of $(x,y)$ for some
integer $n$. If one knows $R$, then the ODE can be solved by quadrature.

We have that:
\begin{equation}
{\frac{\partial (RN)}{\partial x}}+{\frac{\partial (RM)}{\partial y}} = 0.
\label{eq_int_factor}
\end{equation}

\noindent
From~(\ref{eq_int_factor}) we see that
\begin{equation}
  N\, {\frac{\partial R}{\partial x}}
+ R\, {\frac{\partial N}{\partial x}}
+ M\, {\frac{\partial R}{\partial y}}
+ R\, {\frac{\partial M}{\partial y}}
= 0.
\label{eq_int_factor_aberta}
\end{equation}

\noindent
Thus
\begin{equation}
{\frac{D[R]}{R}} =  - \left( {\frac{\partial N}{\partial x}} +
{\frac{\partial M}{\partial y}} \right),
\label{eq_PS}
\end{equation}
where
\begin{equation}
\label{eq_def_D}
D \equiv N {\frac{\partial }{\partial x}}
        + M {\frac{\partial }{\partial y}}
\end{equation}

\noindent
and
\begin{equation}
\label{R}
R = \prod_i f^{n_i}_i
\end{equation}
where $f_i$ are irreducible polynomials and $n_i$ are non-zero
rational numbers. From (\ref{eq_def_D}) and (\ref{R}), we have
\begin{eqnarray}
\label{ratio}
{\frac{D[R]}{R}} & = & {\frac{D[\prod_{i} f^{n_i}_i]}{\prod_i f^{n_k}_k}} =
                    {\frac{\sum_i f^{n_i-1}_i n_i D[f_i] \prod_{j \ne i}
                     f_j^{n_j}}{\prod_k f^{n_k}_k}} \nonumber \\[3mm]
                 & = &  \sum_i {\frac{f^{n_i-1}_i n_i D[f_i]}{f_i^{n_i}}} =
                    \sum_i {\frac{n_iD[f_i]}{f_i}}.
\end{eqnarray}

From~(\ref{eq_PS}), plus the fact that $M$ and $N$ are polynomials,
we conclude that ${D[R]}/{R}$ is a polynomial. Therefore,
from~(\ref{ratio}), we see that $f_i | D[f_i]$ (i.e., $f_i$ is a
divisor of $D[f_i]$).

We now have a criterion for choosing the possible $f_i$ (build all
the possible divisors of $D[f_i]$) and, by using~(\ref{eq_PS})
and~(\ref{ratio}), we have
\begin{equation}
\label{eq_ni}
\sum_i {\frac{n_iD[f_i]}{f_i}} = -\left( {\frac{\partial N}{\partial
x}} + {\frac{\partial M}{\partial y}} \right).
\end{equation}

If we manage to solve~(\ref{eq_ni}) and thereby find $n_i$,
we know the integrating factor for the ODE and the problem is
reduced to a quadrature.

\subsection{Extensions to the Prelle and Singer method}
\label{extensions}

The original PS-procedure deals with rational first order ODEs with
elementary solutions. Extensions were produced, both on the
theoretical basis and on producing new algorithms, in order to
allow tackling more general ODEs.

Using the results presented on \cite{singer}, it is possible to shown
\cite{chris1,firsTHEOps1,secondTHEOps1} that, if the solution for a rational first order ODE of
form (\ref{FOODE}) can be written in terms of Liouvillian functions,
then it presents an integrating factor is of the form:

\begin{equation}
\label{ourR}
R = e^{r_0(x,y)}  \prod_{i=1}^{n} v_i(x,y)^{c_i} =e^{P(x,y)/Q(x,y)}\,S(x,y),
\end{equation}
where $r_0 = P/Q$ ($P$ and $Q$ are polynomials in $(x,y)$) is a rational
function of $(x,y)$, $D[r_0]$ is a
polynomial, $S = \prod_{j} v_i^{\,c_i}$, the $v_i$ are Darboux polynomials (eigenpolynomials) of the $D$
operator and $c_i$ are constants.

Note that this result is the analogous of the previously mentioned
result due to Prelle-Singer but for a more general case, namely, now we
are dealing with Liouvillian solutions (please note that the elementary
solutions, dealt with by Prelle and Singer, are also Liouvillian but are a
restriction to the general Liouvillian case).

In \cite{firsTHEOps1,secondTHEOps1}, we introduced a method that, using
the knowledge embodied by (\ref{ourR}), solves the ODE for the cases where $r_0$ is either
$f(x)$, $g(y)$ or $f(x)+g(y)$, where $f$ and $g$ are rational functions.

In the next section, we will  go a further step by introducing a new method that allows for the
general case for the rational function $r_0$.

\section{Introducing the new Method}

In what follows, we will refine the result (\ref{ourR}) by uncovering
information about the structure of $Q$ ($r_0=P/Q$). This further
knowledge will then be converted into something that will allow us to build
a solving method.

Using (\ref{ourR}) into $D[R]/R=-(\partial_x N + \partial_y M )$, we get:
\begin{equation}
\label{eqadendum1}
D\left[\frac{P}{Q}\right] + \frac{D[S]}{S} =  - \left( \partial_y  M+ \partial_x N \right),
\end{equation}
leading to
\begin{equation}
\label{eqadendum2}
\frac{Q\,D[P]-P\,D[Q]}{Q^2} + \sum_j c_j{\frac{D[v_j]}{v_j}} =   - \left( \partial_y  M+ \partial_x N \right).
\end{equation}

Remembering that the $v_j$ are Darboux polynomials of the $D$ operator, we
can write $D[v_j]=\lambda_j\,v_j$, where the $\lambda_j$ are polynomials in $(x,y)$ (called cofactors) associated with the $v_j$. Thus, we
can write (\ref{eqadendum2}), multiplying both sides by $Q^2$, as:

\begin{equation}
\label{eqadendum4}
Q\,D[P] - P\,D[Q] + Q^2\,\sum_j c_j\,\lambda_j =   - Q^2\,\left( \partial_y  M+ \partial_x N \right).
\end{equation}

The above equation would be the analogous (in our method) to equation (\ref{eq_PS}) (in the PS
method). We call the attention to the fact that we are trying to deal
with a much wider class of ODEs. This is embodied by
the presence of the term $e^{P/Q}$ on the integrating factor. So far, we
only know that $P$ and $Q$ are polynomials thus, it is easy to see that, attempting to
solve (\ref{eqadendum4}), actually means solving a third degree equation for the unknowns: $c_j$ and the coefficients defining the polynomials
$P$ and $Q$. So, our case here seems much poorer than the PS original one as a practical
approach to solving rational first order ODEs (albeit dealing with more general ones).

Can we do something to improve that situation? Actually, we will show
that we can by presenting a refinement to our knowledge about the
general structure for the integrating factor for ODEs of the form (\ref{FOODE}).

\bigskip
\bigskip

{\bf Theorem:} {\it Let the exponent $r_0$ be expressed as $P(x,y)/Q(x,y)$, where $P$
and $Q$ are polynomials in $(x,y)$, with no common factor. Then we have that $Q|D[Q]$ $($i.e., $D[Q]/Q$ is a polynomial
in $(x,y))$.}

\bigskip

{\bf Proof:}
Since $D[r_0]$ is polynomial (see \cite{secondTHEOps1}), we can write it as:

\begin{equation}
\label{EQ_DRzero}
D[r_0] = D\left[\frac{P}{Q}\right] = \frac{Q\,D[P] - P\,D[Q]}{Q^2} = \Pi
\end{equation}
where $\Pi$ is polynomial in $(x,y)$. Multiplying (\ref{EQ_DRzero}) by $Q$, one obtains:

\begin{equation}
\label{EQ_DRzeroQ}
D[P] - \frac{P\,D[Q]}{Q} = \Pi Q.
\end{equation}

Since $D$ is a linear differential operator, with polynomial coefficients, and $P$ is polynomial, $D[P]$ is also polynomial. Therefore, since $\Pi\,Q$ is also polynomial, we may conclude that $\frac{P\,D[Q]}{Q}$ is polynomial either. Since, by hypothesis, $P$ and $Q$ have no common factor, we can infer that $Q|D[Q]$ (i.e., $D[Q]/Q$ is a polynomial), as we wanted to demonstrate
\footnote{An analogous result was demonstrated (by other means) on \cite{chris2}}.

\bigskip
\bigskip
This result, in turn, leads to the following corollary:

\bigskip

{\bf Corollary:} {\it We can write $Q$ as $\prod_{i=1} q_i(x,y)^{m_i}$,
where the $q_i$ are irreducible independent Darboux polynomials of the
$D$ operator and the $m_i$ are positive integers.}

\bigskip
\bigskip

{\bf Proof:} $Q$ is a polynomial so it can be written as $\prod_{i=1} q_i(x,y)^{m_i}$, where the $q_i$ are independent irreducible polynomials and the $m_i$'s are positive integers. So, we can write:

\begin{equation}
\label{DQoverQ}
{\frac{D[Q]}{Q}} =  \sum_{i} m_i{\frac{D[q_i]}{q_i}},
\end{equation}
from the theorem just proved, we know that the left-hand side of
(\ref{DQoverQ}) is polynomial. So the right-hand side is also
polynomial implying that $q_i|D[q_i]$ (i.e., $D[q_i]/q_i$ is a
polynomial), as we wanted to show.

\bigskip
\bigskip

Using these results, we can write $Q = \prod_{i}
\left(v_{\!\hbox{\footnotesize {\it q}}\,i}\right)^{\,m_i}$, where the
$v_{\!\hbox{\footnotesize {\it qi}}}$ are independent irreducible
Darboux polynomials of the $D$ operator, and so:
\begin{equation}
\frac{D[Q]}{Q} = \sum_i\,m_i\,\frac{D[v_{\!\hbox{\footnotesize {\it q}}\,i}]}{v_{\!\hbox{\footnotesize {\it q}}\,i}} = \sum_i\,m_i\,\lambda_{\!\hbox{\footnotesize {\it q}}\,i},
\end{equation}
where the $\lambda_{\!\hbox{\footnotesize {\it qi}}}$ are the
cofactors associated with the $v_{\!\hbox{\footnotesize
{\it qi}}}$. Using this into (\ref{eqadendum4}) and re-arranging we
get:

\begin{equation}
\label{eqadendum5}
D[P] - P\,\sum_i\,m_i\,\lambda_{\!\hbox{\footnotesize {\it q}}\,i} = - \prod_{i} \left(v_{\!\hbox{\footnotesize {\it q}}\,i}\right)^{\,m_i}\,\left(\sum_j c_j\,\lambda_j + \partial_y  M+ \partial_x N \right).
\end{equation}

This equation will prove to be very important to our method. So, let us
now do some analyzing of its structure: What is the advantage of
(\ref{eqadendum5}) in comparison with (\ref{eqadendum4})?

In words, now we know the structure of $Q$, i.e., the building blocks of
$Q$. In (\ref{eqadendum4}), we had to set the degree for $v_i$ (analogously, in
the PS method, we had to set the degree for the $f_i$) and to set the
degree for the $P$ and $Q$ polynomials (there is no analogous in the PS
method). After that, as already pointed out, we are left to solve a
third degree equation on the unknowns. Now, with (\ref{eqadendum5}), we
have to do the same setting of degrees but, due to our improved
knowledge about $Q$, after setting these degrees, we are left with a
finite set of possibilities for the {\bf integers} $\{m_i\}$. Thus
converting solving (\ref{eqadendum5}) to solving  a system of linear equations for
the unknowns $c_j$ and the coefficients defining $P$ (one system of linear equations for each
possible set of integers $\{m_i\}$).

So, after setting the degrees we mentioned, in essence, equation (\ref{eqadendum5})
is as manageable as equation (\ref{eq_PS}) of the PS method. This is the
basis of a new method to deal with rational first order ODEs that can be summarized as
follows:

In equation (\ref{eqadendum5}) (the cornerstone of our method), we see
terms involving the Darboux polynomials (and/or the associated cofactors) of the $D$ operator. So, the first job at hand
is to determine those. In order to do that we have to choose a degree
(as in the PS method). For simplicity sake, let us start with the most
simple possibility, namely, find the Darboux polynomials of degree 1 of
the $D$ operator and the associated cofactors. To be able
to solve equation (\ref{eqadendum5}), we have to set the degree for the $Q$ and
$P$ polynomials. Remembering that we know the structure for $Q$, setting the
degree for $Q$ allows us (see above) to find all possibilities for
$\{m_i\}$. After this, we then try to solve (\ref{eqadendum5}). If we solve it,
we would have found the integrating factor we were looking for. If we
can not solve it, we have to change our settings and try again. The way
we will change the settings\footnote{For example, should we increase the
degree of $P$ without increasing the one for $Q$? Up to which value?
When should we increase the degree for the Darboux polynomials of the $D$
operator? etc.} is a matter of choice and, in principle, it is
difficult to see which one will be the best since this
depends strongly on the ODE itself. The point is that, no matter
which is that choice, the method is contained in the sense that,
even if we have to use high values for the degree for $Q$ and $P$, our task
will be to solve linear equations (sure, it can be a great number of
equations, but always linear ones).

Of course, this procedure can go on indefinitely, but we will be
covering all the possibilities and we may
hope that we will find a solution within our lifetime. On a brighter
tone, all the examples we have come across are solved with degree for
the Darboux polynomials of the $D$ operator, $Q$ and $P$ that allow for a
practical realization of the method.

\section{The Inner Workings of the Method}
\label{examples}

In this section, we are going to present examples of application
of the method. We will present the calculations in a long version
in order for the reader to have a deeper understanding of the
inner workings of the method.

First, in order to illustrate the method just presented, we are
going to start with a simple ODE. This example was chosen due
to its simplicity (so it is a good stating point for the
introducing of the method) and for the fact that, even being
simple, this ODE is not solved by the Maple powerful solver {\tt
dsolve}.

\bigskip
\noindent
{\bf Example 1:}

Consider the following ODE:
\begin{equation}
{\frac {dy}{dx}}={\frac {\left (x+1\right )y}{x-xy -y^{2}+{x}^{2}}}
\end{equation}

For this equation, we have $M=\left (x+1\right )y$ and $N=x-xy -y^{2}+{x}^{2}$ leading to:

\begin{equation}
\label{Dex1}
D = N \partial_x + M \partial_y = (x-xy -y^{2}+{x}^{2}) \partial_x + \left (x+1\right )y \partial_y
\end{equation}
Up to degree 1, we have that the Darboux polynomials (with the associated cofactors) for this operator are:
\begin{itemize}
\item $v_1 = y,\,\,\,\,\,\lambda_1 = x+1$,
\item $v_2 = x+y,\,\,\,\,\,\lambda_2 = 1+x-y$.
\end{itemize}

Then we have to choose the degree for the polynomial $Q$. Starting with
degree 1, since $v_1$ and $v_2$ are of degree 1, the only possible values for
$m_i$ are $\{m_1=1, m_2=0\}$ and $\{m_1=0, m_2=1\}$. Starting
with degree 1 for $P$, we have $P=a_1+a_2\,x+a_3\,y$  and equation (\ref{eqadendum5})
leads to:

\begin{eqnarray}
\label{eqdd0}
{y}^{{\it m1}}\left (x+y\right )^{{\it m2}}\left ({\it n1}\,\left (x+1
\right )+{\it n2}\,\left (1+x-y\right )+3\,x+2-y\right )+\nonumber\\
\left (x-xy-{
y}^{2}+{x}^{2}\right ){\it a2}+\left (x+1\right )y{\it a3}-\nonumber\\
\left ({
\it a1}+{\it a2}\,x+{\it a3}\,y\right )\left ({\it m1}\,\left (x+1
\right )+{\it m2}\,\left (1+x-y\right )\right )
 = 0.
\end{eqnarray}

As we shall see, with the values $m_1=1, m_2=0$ it is possible to find a solution. Substituting  $\{m_1=1, m_2=0\}$ into (\ref{eqdd0}) we get:
\begin{eqnarray}
y\left ({\it n1}\,\left (x+1\right )+{\it n2}\,\left (1+x-y\right )+3
\,x+2-y\right )+\left (x-xy-{y}^{2}+{x}^{2}\right ){\it a2}+\nonumber\\
\left (x+1
\right )y{\it a3}-\left ({\it a1}+{\it a2}\,x+{\it a3}\,y\right )
\left (x+1\right )
 = 0.
\end{eqnarray}

In order to solve the above equation, the coefficients for different
powers of $(x,y)$ have to be zero. Thus leading to the following system of
linear equations:

\begin{eqnarray}
n1+n2+2=0  \nonumber \\
-n2-a2-1=0  \nonumber \\
-a1=0\nonumber \\
n1+n2+3-a2=0
\end{eqnarray}

Leading to the solution for the coefficients:
\begin{equation}
{a1 = 0,\, a3 = a3,\, n1 = 0,\, n2 = -2,\, a2 = 1}
\end{equation}

So, the integrating factor for this ODE becomes (choosing $a_3=0$):
\begin{equation}
R = \frac{e^{x/y}}{\left (x+y\right )^{2}}
\end{equation}

Then the solution is:
\begin{equation}
C=\frac{y\left (-1+y\right ){e^{x/y}}}{\left (x+y\right )}-{e^{-1}}{\it Ei}(1,-{\frac {x+y}{y}})
\end{equation}

We are going to present now an example extracted from the book by Kamke \cite{kamke}. This choice for the second example was made aiming to show
that, even for involved cases, our method is contained and manageable.

\bigskip
\noindent
{\bf Example 2:}

Consider the following ODE ({\bf I.169} from the book by Kamke):
\begin{equation}
\left (ax+b\right )^{2}{\frac {dy}{dx}}+\left (ax+b\right )y^{3}+cy^{2}
\end{equation}

For this equation, we have $M= - \left( (ax+b)y^{3}+cy^{2}\right )
$ and $N=\left (ax+b\right )^{2}$ leading to:

\begin{equation}
\label{Dex1}
D = N \partial_x + M \partial_y = \left (ax+b\right )^{2} \partial_x  - \left( (ax+b)y^{3}+cy^{2}\right ) \partial_y
\end{equation}

For this equation, up to degree 1, we have that the Darboux polynomials (with the associated cofactors) for this operator are:
\begin{itemize}
\item $v_1 = y,\,\,\,\,\,\lambda_1 = -y\,c-b\,y^2-a\,x\,y^2$,
\item $v_2 = (ax+b)/a,\,\,\,\,\,\lambda_2 = a\,b+a^2\,x$.
\end{itemize}

Then we have to choose the degree for the polynomial $Q$. For this
particular example, we will see that we need to use degree 4 for both
$Q$ and $P$. Since $v_1$ and $v_2$ are of degree 1, the possible values
for $m_i$ are $\{m_1=4, m_2=0\}$, $\{m_1=3, m_2=1\}$, $\{m_1=2,
m_2=2\}$, $\{m_1=1, m_2=3\}$ and $\{m_1=0, m_2=4\}$. Letting $P=
a_1\,y^4+a_2\,x^2\,y+a_3\,y^2+a_4\,x\,y^2+a_5\,x\,y+a_6\,x^4+a_7\,
x^3+a_8\,x^2+a_9\,y^3+a_{10}\,x^3\,y+a_{11}\,x\,y^3+a_{12}\,x+a_13\,x^2\,
y^2+a_{14}\,y+a_{15}$, equation (\ref{eqadendum5}) becomes:
\begin{eqnarray}
\label{eqdd}
{y}^{{\it m1}}\left ({\frac {ax+b}{a}}\right )^{{\it m2}}\left ({\it
n1}\,\left (-yc-b{y}^{2}-ax{y}^{2}\right )+{\it n2}\,\left (ab+{a}^{2}
x\right )\right.\nonumber \\
-2\,y\left (axy+by+c\right )\left.-{y}^{2}\left (ax+b\right )+2\,{a
}^{2}x+2\,ab\right )\nonumber \\
+\left ({a}^{2}{x}^{2}+2\,axb+{b}^{2}\right )
\left (2\,{\it a2}\,xy+{\it a4}\,{y}^{2}+{\it a5}\,y+\right.\nonumber \\
\left.4\,{\it a6}\,{x}^
{3}+3\,{\it a7}\,{x}^{2}+2\,{\it a8}\,x+3\,{\it a10}\,{x}^{2}y+{\it
a11}\,{y}^{3}+{\it a12}+2\,{\it a13}\,x{y}^{2}\right )\nonumber \\
-{y}^{2}\left (a
xy+by+c\right )\left (4\,{\it a1}\,{y}^{3}+{\it a2}\,{x}^{2}+2\,{\it
a3}\,y+2\,{\it a4}\,xy+{\it a5}\,x+3\,{\it a9}\,{y}^{2}\right.\nonumber \\
\left.+{\it a10}\,{x}
^{3}+3\,{\it a11}\,x{y}^{2}+2\,{\it a13}\,{x}^{2}y+{\it a14}\right )-
\left ({\it a1}\,{y}^{4}+{\it a2}\,{x}^{2}y+{\it a3}\,{y}^{2}+{\it a4}
\,x{y}^{2}\right.\nonumber \\
\left.+{\it a5}\,xy+{\it a6}\,{x}^{4}+{\it a7}\,{x}^{3}+{\it a8}\,
{x}^{2}+{\it a9}\,{y}^{3}+{\it a10}\,{x}^{3}y+{\it a11}\,x{y}^{3}+{
\it a12}\,x\right.\nonumber \\
\left.+{\it a13}\,{x}^{2}{y}^{2}+{\it a14}\,y+{\it a15}\right )
\left ({\it m1}\,\left (-yc-b{y}^{2}-ax{y}^{2}\right )+{\it m2}\,\left (ab+{a}^{2}x\right )\right) = 0
\end{eqnarray}

As we shall see, with the values $m_1=2, m_2=2$, it is possible to find a solution. Substituting those values for $m_1$ and $m_2$ into (\ref{eqdd}) we get:
\begin{eqnarray}
\label{systemofeqs}
\left \{4\,{b}^{2}{\it a6}+6\,ab{\it a7}+2\,{a}^{2}{\it a8}=0,3\,{b}^{
2}{\it a7}+{a}^{2}{\it a12}+4\,ab{\it a8}=0,\right.\nonumber\\
\left.2\,ab{\it a5}+{\it a12}\,c
+{\it n2}\,{a}^{2}+2\,{b}^{2}{\it a2}+2\,{a}^{2}=0,-3\,c{\it a1}-2\,b{
\it a9}=0,\right.\nonumber\\
\left.{b}^{2}{\it a12}=0,-{\it n1}\,c+{\it a15}\,b+{b}^{2}{\it a4}
-2\,c=0,2\,{a}^{2}{\it a2}+{\it a7}\,c+6\,ab{\it a10}=0,\right.\nonumber\\
\left.8\,ab{\it a6}+
3\,{a}^{2}{\it a7}=0,2\,{b}^{2}{\it a8}+2\,ab{\it a12}=0,4\,{a}^{2}{
\it a6}=0,\right.\nonumber\\
\left.-3\,b-{\it n1}\,b+{b}^{2}{\it a11}-c{\it a3}=0,{a}^{2}{\it
a5}+3\,{b}^{2}{\it a10}+{\it a8}\,c+4\,ab{\it a2}=0,\right.\nonumber\\
\left.b{\it a12}+{\it
a15}\,a+2\,{b}^{2}{\it a13}+2\,ab{\it a4}=0,-{\it n1}\,a-3\,a-c{\it a4
}+2\,ab{\it a11}=0,\right.\nonumber\\
\left.b{\it a8}+{a}^{2}{\it a4}+a{\it a12}+4\,ab{\it a13}
=0,{\it n2}\,ab+2\,ab+{b}^{2}{\it a5}+{\it a15}\,c=0,\right.\nonumber\\
\left.{a}^{2}{\it a11}-
c{\it a13}=0,-b{\it a3}-2\,c{\it a9}=0,{\it a6}\,a=0,3\,{a}^{2}{\it
a10}+{\it a6}\,c=0,\right.\nonumber\\
\left.-2\,b{\it a11}-2\,a{\it a9}=0,-3\,a{\it a1}=0,-3\,b
{\it a1}=0,-a{\it a13}=0,-2\,a{\it a11}=0,\right.\nonumber\\
\left.-b{\it a13}-a{\it a4}=0,-2\,
c{\it a11}-a{\it a3}-b{\it a4}=0,\right.\nonumber\\
\left.a{\it a8}+b{\it a7}+2\,{a}^{2}{\it
a13}=0,a{\it a7}+b{\it a6}=0\right \}
\end{eqnarray}

Leading to the solution for the coefficients:
\begin{eqnarray}
\left \{{\it a6}=0,{\it a11}=0,{\it a10}=0,{\it a1}=0,{\it n2}=-1,{
\it a2}=0,{\it a7}=0,{\it a9}=0,\right.\nonumber\\
\left.{\it a3}=-1/2\,{\frac {{c}^{2}-2\,a{b}
^{2}{\it a13}}{{a}^{3}}},{\it n1}=-3,{\it a14}=-{\frac {cb}{{a}^{2}}},\right.\nonumber\\
\left.{\it a15}=-1/2\,{\frac {{b}^{2}}{a}},{\it a5}=-{\frac {c}{a}},{\it a12
}=-b,{\it a4}=2\,\right.\nonumber\\
\left.{\frac {b{\it a13}}{a}},{\it a13}={\it a13},{\it a8}=
-a/2\right \}
\end{eqnarray}

So, the integrating factor for this ODE becomes (choosing $a_{13}=0$):
\begin{equation}
R = \frac{{e^{-\,{\frac {\left (yc+{a}^{2}x+ab\right )^{2}}{2\,a{y}^{2}\left (ax
+b\right )^{2}}}}}}{{y}^{3}\left (ax+b\right )}
\end{equation}

Then the solution is:
\begin{equation}
C = \int \frac{\left (axy+by+c\right ){e^{\frac{-1}{2\,a}\,\left ({\frac {c}{
ax+b}}+{\frac {a}{y}}\right )^{2}}}}{{y}\left (ax+b\right )}
{dx}
\end{equation}

From the example above, we can see that, even in cases where $Q$ and $P$ have high
degrees, our method convert the problem into a simple linear algebraic
system. More specifically, in this example, the equations
(\ref{systemofeqs}). Although these equations look frightening, they are
solvable even by hand (if one is brave enough). In this computational
implementation, the problem is trivially solved.

Let us then introduce our package implementing the above
explained ideas and making the algorithm available for easy usage.

\section{The {\it Lsolver} package}
\label{package}

\subsection*{\it Summary}

A brief review of the commands of the package is as
follows:\footnote{This subsection and the next one may contain some
information already presented in the previous sections; this is
necessary to produce a self-contained description of the package.}

\begin{itemize}

\item {\tt Lsolve} solves ODEs, using our Darboux-type approach. It deals only with rational first order ODEs.

\item {\tt IntFact}  returns an integrating factor for the rational first order ODE.

\item {\tt LDop} constructs the ${D}$ operator associated to the rational first order ODE.

\item {\tt Darboux} returns the DPs (Darboux polynomials) and cofactors associated with the  ${D}$  operator.

\end{itemize}

\subsection*{\it Description}

A complete description of the {\it Lsolver} package's commands can also
be found in the on-line help.

\subsection{Command name: {\tt Lsolve}}
\label{Lsolve}

\noindent {\it Feature:} This command applies our Darboux-type procedure to solve ODEs.

\bigskip

\noindent
{\it Calling sequence\footnote{In what follows, the {\it input}
can be recognized by the Maple prompt \verb->-.}:}
\begin{verbatim}
> Lsolve(ODE,extra_args);
\end{verbatim}

\noindent
{\it Parameters:}\newline

\noindent
\begin{tabular}{ll}
\verb-ODE-         & - a first order ordinary differential equation.
\\&
\\

\verb-extra_args-     & {\tt- Deg = <bound>} - a positive integer equal to the maximum degree of \\ & the
 darboux polynomials. The default value is 1.\\&\\ &   {\tt - Deg\_Q = <bound\_Q>} - a positive integer  equal to the maximum degree of\\ &the Q polynomial; The default value is 5.\\&\\
&  {\tt - Deg\_P = <bound\_P>} - a positive integer equal to the maximum degree\\ & of the P polynomial; The default value is 5.\\

\end{tabular}

\bigskip

\noindent
{\it Synopsis:}
\smallskip
\smallskip

The {\tt Lsolve} command is a part of the {\it Lsolver} package, which is designed to solve rational first order ODEs.

\medskip
\noindent
{\it The arguments}
\smallskip
\smallskip

The first argument of {\tt Lsolve} is the ODE. The extra arguments are all positive integers that specify the polynomial degrees we are going to use in the
search for the integrating factor. We would like to remind the reader that our Darboux-type procedure is a semi-decision one, in other words, if the integrating factor
has not be found for these degrees, it does not exists at this level. But, it could
as well exist for higher degree, we have to keep looking.

\bigskip

\noindent

\noindent
\subsection{Command name: {\tt IntFact}}
\label{IntFact}

\noindent
{\it Feature:} Calculates an integrating factor for the ODE.
\smallskip
\smallskip
\smallskip

\noindent {\it Calling sequence:}
\begin{verbatim}
> IntFact(ODE,extra_args);
\end{verbatim}

\noindent
{\it Parameters:}
\smallskip
\smallskip

\noindent
The parameters {\tt ODE} and {\tt extra\_args} have the same meaning as explained
above\footnote{In what follows, these parameters will always have this meaning.}.
The command {\tt IntFact} admits all {\tt the extra\_args} as the command {\tt Lsolve}.

\bigskip

\smallskip
\smallskip

\medskip
\noindent
{\it Synopsis:}
\smallskip
\smallskip

Actually this command embodies the heart of the package since, after
finding the integrating factor, the solution is found via a quadrature.

\noindent
\subsection{Command name: {\tt Darboux}}
\label{IntFact}

\noindent
{\it Feature:} Calculates the Darboux polynomials, up to a certain
degree, of the $D$ operator related to the ODE in question. The
default is to use degree=1 for the Darboux polynomials to be sought.
\smallskip
\smallskip
\smallskip

\noindent {\it Calling sequence:}
\begin{verbatim}
> Darboux(ODE,extra_args);
\end{verbatim}

\noindent
{\it Parameters:}
\smallskip
\smallskip

\noindent
The command {\tt Darboux} admits as {\tt extra\_args} only {\tt Deg =
<bound>}, where {\tt <bound>} is a positive integer equal to the maximum degree of the darboux
polynomials.

\bigskip

\smallskip
\smallskip

\medskip
\noindent
{\it Synopsis:}
\smallskip
\smallskip

This command is very important to the method since it determines the
building blocks for the integrating factor.

\noindent
\subsection{Command name: {\tt LDop}}
\label{Dop}

\noindent {\it Feature:} Returns the ${D}$ operator (for the ODE) that is
an essential ingredient in applying our procedure.

\smallskip
\smallskip
\smallskip

\noindent {\it Calling sequence:}
\begin{verbatim}
> Dop(ODE);
\end{verbatim}

\noindent
{\it Synopsis:}
\smallskip
\smallskip

This command returns the ${D}$ operator associated with the ODE.

\subsection{Example of the usage of the package commands}
\label{simples}
\indent

Let us use some rational first order ODEs to show the commands in action. Consider the following ODE:

\begin{equation}
{\frac{{\rm d}y}{{\rm d}x}}=\frac{y(1+x)}{x+x^2-y^2}.  \label{simple}
\end{equation}

Usually, people only want to find the solution for the ODE. So, that would
simply require the input lines:

\begin{verbatim}
> eq := diff(y(x),x) = y(x)*(1+x)/(x+x^2-y(x)^2);
> Lsolve(eq);
                     The solution will be
\end{verbatim}
\begin{equation}
-i\sqrt {\pi }\sqrt {2}{\it erf} \left( {\frac {1/2\,i\sqrt {2}x}{y
 \left( x \right) }} \right) +2\,{e^{1/2\,{\frac {{x}^{2}}{ \left( y
 \left( x \right)  \right) ^{2}}}}}y \left( x \right) ={\it \_C1}
\end{equation}

Sometimes the user is interested on having a look at just the integrating factor, so
he/she could type:
\begin{verbatim}
> IntFact(eq);
                   For the ODE in the form
\end{verbatim}
\begin{equation}
{\frac{{\rm d}y}{{\rm d}x}}=\frac{y(1+x)}{x+x^2-y^2}
\end{equation}
\begin{verbatim}
                 the integrating factor will be
\end{verbatim}
\begin{equation}
\frac{\exp\left(\frac{x^2}{2y^2}\right)}{y^2}
\end{equation}

\noindent
Please notice that the program informs the user the format in which it is
considering the ODE to be in, that is important since it may be the case where
the input ODE gets transformed by the simplification routines and is regarded
in a different shape (of course it is still the same ODE) and that will affect the
integrating factor.

If one wants to find the Darboux polynomials (up to a certain degree),
one can use the following input:

\begin{verbatim}
> Darboux(eq);

    Below, you can find a list containing two lists:
    [[Darboux polynomials],[co-factors]]

\end{verbatim}
\begin{equation}
[[y],[1+x]]
\end{equation}

\noindent
Please remember that the default value for the degree of the
Darboux polynomials is $1$.

Finally, for this example, let us calculate the $D$ operator:

\begin{verbatim}
> LDop(eq):
\end{verbatim}

\begin{equation}
w \rightarrow (x+x^2-y^2)(\frac{d}{dx}\,\,w)+y(1+x)(\frac{d}{dy}\,\,w)
\end{equation}

Let us deal with a more involved case. Consider the following ODE.

\begin{equation}
{\frac {dy}{dx}} ={\frac {-1+x+y +3\,y^{2}}{2(2\,x+y +xy+ y^{2
}- y^{3})}}
\end{equation}

Again, using the input lines just
introduced above:

\begin{verbatim}
> eq := diff(y(x),x) = 1/2*(-1+x+y(x)+3*y(x)^2)/(2*x+y(x)+x*y(x)+
  y(x)^2-y(x)^3);
> Lsolve(eq);
                     We could not find an integrating factor
\end{verbatim}

What happened? The reason for the failure in solving the equation above
could be due to the fact that there is no Darboux polynomials with degree equals
to $1$ and, since the default for the command assumes that value, we
have to increase this value to check that possibility. This is done via the following input line:

\begin{verbatim}
> Lsolve(eq,Deg=2);

                     The solution will be

\end{verbatim}
\begin{equation}
2\,\sqrt {x+ \left( y \left( x \right)  \right) ^{2}}{e^{1/4\,{\frac {
-1+4\,x+4\,y \left( x \right) }{x+ \left( y \left( x \right)  \right) 
^{2}}}}}-3\,\sqrt {\pi }{e^{1}}{\it erf} \left( 1/2\,{\frac {-1+2\,y
 \left( x \right) }{\sqrt {x+ \left( y \left( x \right)  \right) ^{2}}
}} \right) ={\it \_C1}
\end{equation}

Let us have a look at the other commands:

\begin{verbatim}
> IntFact(eq,Deg=2);
                   For the ODE in the form
\end{verbatim}
\begin{equation}
{\frac {{\it dy}}{{\it dx}}}={\frac {-1+x+y+3\,{y}^{2}}{4\,x+2\,y+2\,y
x+2\,{y}^{2}-2\,{y}^{3}}}
\end{equation}
\begin{verbatim}
                 the integrating factor will be
\end{verbatim}
\begin{equation}
{e^{{\frac {-1/4+x+y}{x+{y}^{2}}}}} \left( x+{y}^{2} \right) ^{-3/2}
\end{equation}

The other two commands, for this case, produce:

\begin{verbatim}
> Darboux(eq,Deg=2);

    Below, you can find a list containing two lists:
    [[Darboux polynomials],[co-factors]]

\end{verbatim}
\begin{equation}
[[x+{y}^{2}],[4+4\,y]]
\end{equation}
and
\begin{verbatim}
> LDop(eq):
\end{verbatim}

\begin{equation}
w \rightarrow ((4x+2y+2yx+2y^2-2y^3))(\frac{d}{dx}\,\,w)+(-1+x+y+3y^2)(\frac{d}{dy}\,\,w)
\end{equation}

\section{Performance}
\label{performance}
\indent

As previously mentioned, we have used rational ODEs to exemplify
the usage of the package commands, we hope to have clarified that
usage. Now we are going to further exemplify the package but with
a different approach. We will not be concerned with the
input/output of the commands. In order to analyze the contribution
of the present implementation, we will comment on how the package
extends the solving capabilities of the existing solvers and
how it does that. We will divide that into two main groups and
present examples of those:

\subsection{\it First order ODEs with Liouvillian solutions}
\label{lis}

The first one is the most obvious. The present implementation
tackles (semi-algorithmically) rational first order ODEs with Liouvillian solutions
and those were not treated neither on the original PS-approach nor
in its extensions. So, this implementation extends the
applicability of these Darboux-type approaches to an wider range
(not only on the Maple environment).

An example of a rational first order ODE belonging to this class is:

\begin{equation}
{\frac{{\rm d}y}{{\rm d}x}}= -{\frac{y^2(-2y+1-2x+x^2y)}{x^2(-2x+1-2y+xy^2)}}
\end{equation}

{\it Lsolver} finds the following integrating factors for this ODE:

\[{\frac{\exp\left({\frac{2x^2+y^2}{x^2y^2}}\right)}{x^4y^4}}\]
\noindent
which, in turn, lead to the following solution:
\[
\int\frac{\exp\left({\frac{(x+y)^2}{x^2y^2}}\right)(-2y+1-2x+x^2y)}{x^2}dx = C
\]

\bigskip
As previously mentioned, this equation would escape the standard PS
procedure. Furthermore, this example (which is  just an example of an infinite class) also elude the {\tt dsolve}
command.

We would like to point out that the Maple solver can deal with some
members of this class, for instance the Abel's ODEs but not in the
general case. Our previous implementation extending the Prelle-Singer method
\cite{nosso_CPC_PS1} also solve some ODEs but with heuristic approaches.

\subsection{\it First order ODEs with elementary solutions}
\label{elfs}

The second group talks about a bonus: The present implementation
solves ODEs, with elementary solutions, that are missed by the
powerful solvers on the Maple commercial algebraic package and in
some of the (Maple) implementations of the PS-approach and its
extensions already mentioned. We call that a bonus
since the present package was designed to dwell in the previously
uncharted region (at least semi-algorithmically) of the rational first order
ODEs with Liouvillian solutions but it fared well elsewhere
also, as we shall see. But, before we go on with the example, let us
answer the following: How can an implementation of the PS-approach
\cite{nosso_CPC_PS1} not solve ODEs with elementary solutions as we
are claiming here?

Of course, theoretically, the PS-approach would eventually catch
the integrating factor, given enough time and computational power.
Here it is where the situation gets complicated (at least for
ODEs like the one we are about to display). In practice, the
Darboux polynomials can be of a too high a degree and that
puts matters outside the nowadays computational capabilities (both on time and in computer memory). The
attentive reader might ask: Why does this practical limitation not
affect the extension here presented? The point is that the general
form of the Integrating factor (\ref{ourR}) can allow for a solution
with lower Darboux polynomial degree to be found and that would turn the solving of the ODE feasible. Let us
present then an example of these ODEs to clarify these ideas:

$${\frac {{\it dy}}{{\it dx}}}=(-14\,x-14\,y-28\,{x}^{3}+14\,{y}^
{3}+40\,{x}^{4}-58\,{x}^{5}-19\,{x}^{2}y+30\,{x}^{3}y+$$

$$-23\,{x}^{2}{y}^{
2}+26\,{x}^{3}{y}^{2}+14\,x{y}^{3}+21\,{x}^{4}y)\,{\Large /}\,(x \,( 7\,{x}^{2}+7
\,{x}^{3}+7\,x+$$

\begin{equation}
\label{eqex}
+7\,y+7\,xy+7\,{y}^{2}+13\,{x}^{2}y+7\,x{y}^{2}+13\,{x}^
{3}y+7\,{x}^{4} ) )
\end{equation}

Our program finds the following integrating factor for the above ODE:
\begin{equation}
\label{Rex}
R = {e^{{\frac {y-1}{{x}^{2}}}}} \left( x+1 \right) ^{-8}
\end{equation}

which, in turn, lead to the following solution:
\begin{equation}
\label{sex}
C=\frac{{e^{{\frac {y-1}{{x}^{2}}}}}{x}^{3} \left( 7\,x+7
\,{x}^{3}+{x}^{4}-{x}^{2}y - 7\,{x}^{2}+7\, y^{2}+7\,y  \right)}{
 \left( 21\,{x}^{5}+{x}^{7}+7\,{x}^{6}+35\,{x}^{4}+35\,{x}^{3}+21\,{x}
^{2}+7\,x+1 \right)}
\end{equation}

From (\ref{sex}) and (\ref{Rex}), one can see that (\ref{eqex}) admits
an integrating factor ${\overline R}$ of the form:

\begin{equation}
\label{Rex2}
{\overline R} = {\frac {21\,{x}^{5}+{x}^{7}+7\,{x}^{6}+35\,{x}^{4}+35\,{x}^{3}+21\,{x}
^{2}+7\,x+1}{ \left( x+1 \right) ^{8}{x}^{3} \left( 7\,x+7\,{x}^{3}+{x
}^{4}-{x}^{2}y-7\,{x}^{2}+7\,{y}^{2}+7\,y \right) }}
\end{equation}

So, in principle, the PS-approach should find it. As we pointed
out above, in practice this does not occur since (please see
(\ref{Rex2})) one would need to find Darboux polynomials up to
degree 7. In our case, please look at (\ref{Rex}), we needed to
use Darboux polynomials of degree 1.

We remind the reader that the {\tt dsolve} command contains many
heuristic methods based on Lie theory for solving ordinary
differential equations \cite{nosso,nosso2}. But, because of their
pattern-matching based nature, there will always be an infinity of
cases (of which the equation above is an example), for which the
symmetries can not be found.

\subsection{\it Time Performance}

In order to provide information regarding the computational time
performance of the package in solving ODEs, we present the table
\ref{tabela} with the ODEs we have analysed in the paper. The
reason for this choice is that, no matter how large the sample we
chose to display, it will never be exhaustive, there are an
infinity of possibilities. So, we have decided to use the ODES
to which the readers have more familiarity (since they have seen
the equations being discussed) in order for them to better analyze
which feature influences the most the time of solution. Only one
ODE analyzed in the paper was left out of the table, equation
(\ref{eqex}). That is so, since it is too big for it to fit the
table in a reasonable way. The time the program took to solve this
ODE was $1.713\,s$ (please see the data related to the other
ODEs on the table). As can be seen from the comparison of these
data, the time the program took to solve this ODE is not the
biggest of them all, even the ODE itself being the ``scariest''.
This shows that not only the size of the ODE matters to the time
of solving it. All these times were obtained on a machine with an
Intel Celeron M processor 360 (1.40 GHz), 256 MB DDR SDRAM,
running Windows XP Home.

\begin{table}[h]
\centering
\begin{tabular}{|l|l|l|}
\hline
ODE & Solution & Time \\
\hline
${\frac {dy}{dx}} ={\frac {y  \left(
1+x \right) }{x-xy - y^{2}+{x}^{2}}}$
& $ \left( {\it C}+{\it Ei} \left( 1,-{\frac {x+y }{y}} \right) \right) \,( y + x )+\left( y - y^{2}
 \right) {e^{{\frac {x+y}{y}}}}=0$ & $0.380s$ \\
\hline
$ {\frac {dy}{dx}} =-{\frac {y^{2} \left( y ax+yb+c \right) }{{a}^{2}{x}^{2}+2\,axb+{b}^{2}}}
$& $\int \!{e^{-{\frac { \left( ab+cy+{a}^{2}x \right) ^{2}
}{2a{y}^{2} \left( ax+b \right) ^{2}}}}} \frac{\left( yax+yb+c \right)}{
 \left( ax+b \right)}{dx}+{\it C}\,y=0
$ & $4.096s$ \\
\hline

${\frac {dy}{dx}}={\frac {y\left( 1+x \right) }{x- y^{2}+{x}^{2}}}
$ & $i\sqrt{\frac{\pi}{2}}{\it erf} \left( {\frac {i\sqrt {2}x}{2y}} \right) -{e^{{\frac {{x}^{2}}{2y^{2}}}}}y +{\it C}=0
$  & $0.550s$ \\
\hline

${\frac {dy}{dx}} ={\frac {-1+x+y+3\, y^{2}}{2(2\,x+y+xy +y^{2
}-y^{3})}}
$ & $\sqrt {x+y^{2}}{e^{-{\frac
{ \left( -1+2\,y  \right) ^{2}}{4(x+y^{2})}}}}-\frac{3}{2}\sqrt {\pi }{\it erf} \left( {
\frac {-1+2\,y }{2\sqrt {x+y^{2}}}} \right) -{\it C}=0
$ & $1.002s$ \\
\hline

${\frac {dy}{dx}}=-{\frac {y^{2} \left( -2\,y+1-2\,x+{x}^{2}y
\right) }{{x}^{2} \left( -2\,x+1-2\,y+y^{2}x \right) }}
$ & $\int \!{e^{{\frac { \left( x+y \right) ^{2}}{{y}^{2}{x}^{2}}}}}
 \frac{\left( -2\,y+1-2\,x+{x}^{2}y \right)}{{x}^{2}}{dx}+{\it C}=0
$ & $1.562s$\\
\hline

\end{tabular}

\caption{ODEs analyzed in the paper, their solutions and the corresponding time
the program takes to solve them.} \label{tabela}
\end{table}

\section{Conclusions}
\label{conclude}

The Prelle-Singer approach is a semi-decision algorithm that
can be used to solve analytically rational first order ordinary
differential equations. Some implementations of it have been
produced \cite{Man1,nosso_CPC_PS1}. In \cite{nosso_CPC_PS1}, we
have introduced an extension to the PS-approach and we have
presented an implementation of a package that extends the
applicability of the PS-type methods in order to tackle some
rational first order ordinary differential equations whose
solutions contain a sub-class of (non-elementary) Liouvillian
functions (therefore, outside the scope of the original
PS-approach).

Here, we have implemented an extension presented on \cite{JCAM}. This
extension now covers (in the same semi-decision way as the original PS
approach does but for a wider class of differential equations) all
rational first order ordinary differential equations with Liouvillian
solutions. So, this enlarges the scope of the PS-type procedures
previously implemented \cite{Man1,nosso_CPC_PS1} to allow it to tackle
rational first order ODEs with Liouvillian solutions (see subsection \ref{lis}).
But, furthermore, it is worth mentioning that, as an additional bonus to
this, we have enhanced the ODE solving capabilities in Maple, even for
the cases where the ODE presents elementary solutions (a sub set of
the Liouvillian ones). Since, in the present implementation, the
integrating factor has a different general form: it is written with an
exponential part and, in some cases, that translates to computational
speed in the determination of it since the integrating factor can be
found with Darboux polynomials of much lower degree and that can be the
difference between failure and success in solving the rational first
order ordinary differential equations (see subsection \ref{elfs}).

In other words, apart from the theoretical extension that the
present package now makes available for the user, thus making the
analytical solving of a new class of rational first order ordinary
differential equations possible, it also improves the Maple
solving capabilities in the arena where the {\tt dsolve} Maple
command (even with some of its non-default enhancements turned on,
e.g. the Lie package) and some previously made implementations of
the PS-approach and its extensions \cite{Man1,nosso_CPC_PS1}
already act.

As further development for this work, we are currently working on extensions to
the package to tackle ODEs containing elementary functions (in the same
fashion Shtokhamer \cite{Shtokhamer} used for the usual PS procedure).

\end{document}